\documentclass[10pt]{article}
\usepackage[OE]{express}
\usepackage{boldline}

\begin{document}
\title{Metal-Dielectric Antennas for Efficient Photon Collection from Diamond Color Centers}

\author{Amir Karamlou$^{1,2,\dagger,+}$, Matthew E. Trusheim$^{1,\dagger,\times}$, and Dirk Englund$^{1,*}$}

\address{$^1$Department of Electrical Engineering and Computer Science, Massachusetts Institute of Technology, Cambridge, Massachusetts 02139, USA\\
	$^2$Department of Physics, Massachusetts Institute of Technology, Cambridge, Massachusetts 02139, USA\\
	$^\dagger$ Equal contribution
}

\email{$^+$karamlou@mit.edu}
\email{$^\times$mtrush@mit.edu}
\email{$^*$englund@mit.edu} %% email address is required



\begin{abstract}
	A central challenge in quantum technologies based on atom-like defects is the efficient collection of the emitter's fluorescence. Optical antennas are appealing as they offer directional emission together with spontaneous emission rate enhancement across a broad emitter spectrum. In this work, we introduce and optimize metal-dielectric nanoantenna designs recessed into a diamond substrate and aligned with quantum emitters. We analyze tradeoffs between external quantum efficiency, collection efficiency, radiative Purcell factor, and overall collected photon rate. This analysis shows that an optimized metal-dielectric hybrid structure can increase the collected photon rate from a nitrogen vacancy center by over two orders of magnitude compared to a bare emitter. As a result, these metal-dielectric antennas should enable single-shot electron spin measurements of NV centers at room temperature.
\end{abstract}

\ocis{(270.0270) Quantum optics; (270.5585) Quantum information and processing}% REPLACE WITH CORRECT OCIS CODES FOR YOUR ARTICLE, MINIMUM OF TWO; Avoid using the OCIS codes for “General” or “General science” whenever possible.
%For a complete list of OCIS codes, visit: https://www.osapublishing.org/oe/submit/ocis/

%%%%%%%%%%%%%%%%%%%%%%% References %%%%%%%%%%%%%%%%%%%%%%%%%
\bibliography{main} 
\bibliographystyle{osajnl}

%%%%%%%%%%%%%%%%%%%%%%%%%%  body  %%%%%%%%%%%%%%%%%%%%%%%%%%
\section{Introduction}
Optical antennas have been widely explored for engineering strong interactions between light and quantum emitters. Nitrogen-vacancy (NV) centers in diamond have been a particular focus due to their spin coherence properties and applications in quantum sensing and information processing \cite{Jelezko2006}. A central problem for many of these applications concerns the overall fluorescence collection rate, which is correlated to the readout fidelity of the NV electron spin system and consequently impacts NV applications in sensing and quantum information processing\cite{Degen2017}. The collected photon rate (CPR) increases with the photon emission rate, proportional to the Purcell factor $F_p$, and the photon collection efficiency (CE), which depends on the spatial overlap of the emission with the collection optics. A variety of different approaches have been taken to increase these factors independently\cite{Novotny2011,Chroder2016}. Dielectric structures have been shown to achieve high CE across a broad spectrum\cite{Li2015a} or narrow-spectrum Purcell enhancement\cite{Li2015} while metallic structures have achieved broadband radiative enhancement\cite{Waks2010} but low collection efficiency\cite{Choy2011}. In this work, we do not attempt to optimize $F_P$ or CE individually, but rather the overall enhancement in photon collection rate, CPR $\propto F_P \cdot $CE. As we will show, maximizing CPR requires a trade-off between CE and $F_P$. We consider a number of metal-dielectric antenna geometries, including all-dielectric, metallic, and hybrid metal-dielectric designs. All structures provide an increase in CPR as compared to bare emitters, with each offering a different trade-off between CE and $F_P$. Integrating across the NV spectrum, a metal-dielectric optical antenna achieves the highest CPR enhancement of 25.6, which represents a 400-fold increase over an NV in unmodified diamond.

%In particular, plasmonic resonators have achieved high Purcell enhancements when coupled to variety of emitters including organic molecules, quantum dots, and color centers in diamond \cite{moerner, loncar1, benson1, mikkelsen1}. Plasmonic bowtie antennas coupled to nanodiamonds containing nitrogen vacancy (NV) centers have been shown to achieve broadband Purcell enhancement, reducing the NV lifetime by as much as a factor of nine \cite{benson2}, while increasing collected fluorescence by a similar amount. Metallic apertures fabricated in bulk diamond \cite{loncar2} have also been shown to reduce the NV excited state lifetime by up to six-fold and additionally yielded an overall improvement in collected fluorescence flux by a factor of twelve compared to NV centers in unpatterned diamond substrates.Emitter-cavity systems are well-described in the language of cavity QED, where the loss rates $\kappa$, $\gamma$ and coupling rate g characterize the system \cite{cavityQEDreview}.Collected photon rate enhancement is in the context of color centers in diamond, such as the NV$^-$ and silicon vacancy (SiV). 

Central to room-temperature NV applications is the ability to measure its electronic spin state optically by collecting its fluorescence emission, a technique known as optically-detected magnetic resonance\cite{Doherty2013a}. Under off-resonant pumping, the NV cycles between its ground $^3$A and excited $^3$E states, leading to fluorescence emission across a broad spectrum consisting of a zero-phonon line peak at 637 nm and a phonon side band with a maximum intensity at around 680 nm (Fig. 1). This emitted fluorescence is correlated with the NV electron spin state, as the m$_s$ = $\pm 1$ states couple non-radiatively to the singlet $^1$A levels, reducing the fluorescence emission rate of these states in the visible range. In turn, these singlet levels decay non-radiatively (lifetime $\sim$ 300 ns) and preferentially to the m$_s$ = 0 spin ground state, resulting in spin polarization \cite{Doherty2013a}. A $> 90\%$ polarization of individual NV centers as well as efficient spin readout, is thus possible at room temperature, enabling the many applications of NV centers. The SNR of optical spin readout, which translates directly into sensitivity and quantum fidelity, is limited by the shot noise on the detected photon number as well as the intensity contrast between the spin states. At the current state of the art \cite{Li2015a,Shields2015}, the number of fluorescence photons detected in the readout window (i.e. before the spin becomes re-polarized, destroying information about its state) is approximately 3, corresponding to a steady-state count rate of $\sim$ 1 Mcps. Although more advanced methods can be used to increase spin state contrast, including spin-to-charge conversion \cite{Shields2015} or repetitive readout using nuclear spin ancilla \cite{2009.Science.Jiang-Lukin.repetitive_readout_spin_NV}, the photon shot noise on this measurement precludes the bare NV electronic system from reaching the spin-projection noise limit at room temperature\cite{Wolf2015}. The spin readout of other diamond color centers such as the silicon- or germanium-vacancy, although currently performed at low temperature unlike the NV, is similarly limited by the rate of detected fluorescence photons\cite{Muller2014a,Rogers2014b,Siyushev2016}.

To reach the fundamental noise limit in optical spin readout, a higher fluorescence detection rate is needed. Assuming that the NV spin-mixing process is not affected by Purcell enhancement, the relevant figure of merit is a product of the Purcell factor and CE\cite{Wolf2015}. Crucially, this figure of merit must be optimized across the wide fluorescence spectrum of the NV center, which at room temperature spans hundreds of nanometers. Broadband nanoantennas (Fig. 1) are a natural choice. Because the mode volume of a nanoantenna can be extremely small, the Purcell factor\cite{DeLeon2012a}

\begin{equation}
F_p=\frac{\Gamma}{\Gamma_{free}}=\frac{3\lambda^3Q}{4\pi^2V_{eff}}
\end{equation}
can be large even while the $Q$-factor remains small, allowing for broadband spontaneous emission enhancement. Here, $\Gamma$ is the dipole emission rate in the presence of the nanoantenna, $\Gamma_{free}$ is the dipole emission rate in free space, and $V_{eff}$ is the effective mode volume of the nanoantenna. In addition, nanoantennas can shape the far-field emission of the optical emitter dipole and create a directional pattern that maximizes power within a given numerical aperture (NA). The resulting CE \cite{Agio2013} is given by
\begin{equation}
CE=\frac{1}{2Z_0}\frac{\int\int \left| E(\theta,\phi) \right|^2 \sin(\theta) d\phi d\theta}{P_{out}},
\end{equation}

where $P_{out}$ is the power emitted by the dipole, $Z_0$ is the impedance of the surrounding medium, $|E|^2$ is the electric field intensity in the far-field, and $0 < \theta < \sin^{-1}(NA)$ is the collection angle associated with a given numerical aperture. Although non-radiative processes or coupling to local modes would increase the Purcell factor (increase $P_{out}$), emission into these modes does not reach the far-field and results in a lower CE. In this way, the CE corrects for any reduction in external quantum efficiency due to loss that occurs in the optical structures \cite{Thomas2004}, as well as losses due to total internal reflection at the diamond-air interface. The enhancement in the collected fluorescence rate into a given microscope collection NA from a single quantum emitter due the presence of an optical antenna can therefore be quantified by the figure of merit $CPR = F_p \times CE$. A structure that results in perfect collection of emitted fluorescence (CE = 1) but no Purcell enhancement ($F_p$ = 1) would achieve a CPR of 1. A typical CPR for an emitter embedded in a high-index substrate using oil-immersion microscopy is $\sim 0.05$, as total internal reflection limits CE. In this work, we seek to improve on the state of the art by introducing and numerically optimize novel metallic, dielectric and hybrid designs that are able to increase CPR across the broad NV spectrum.

\begin{figure}[h!]
	
	\begin{minipage}[b]{0.5\linewidth}
		\centering
		\includegraphics[width=13.2cm]{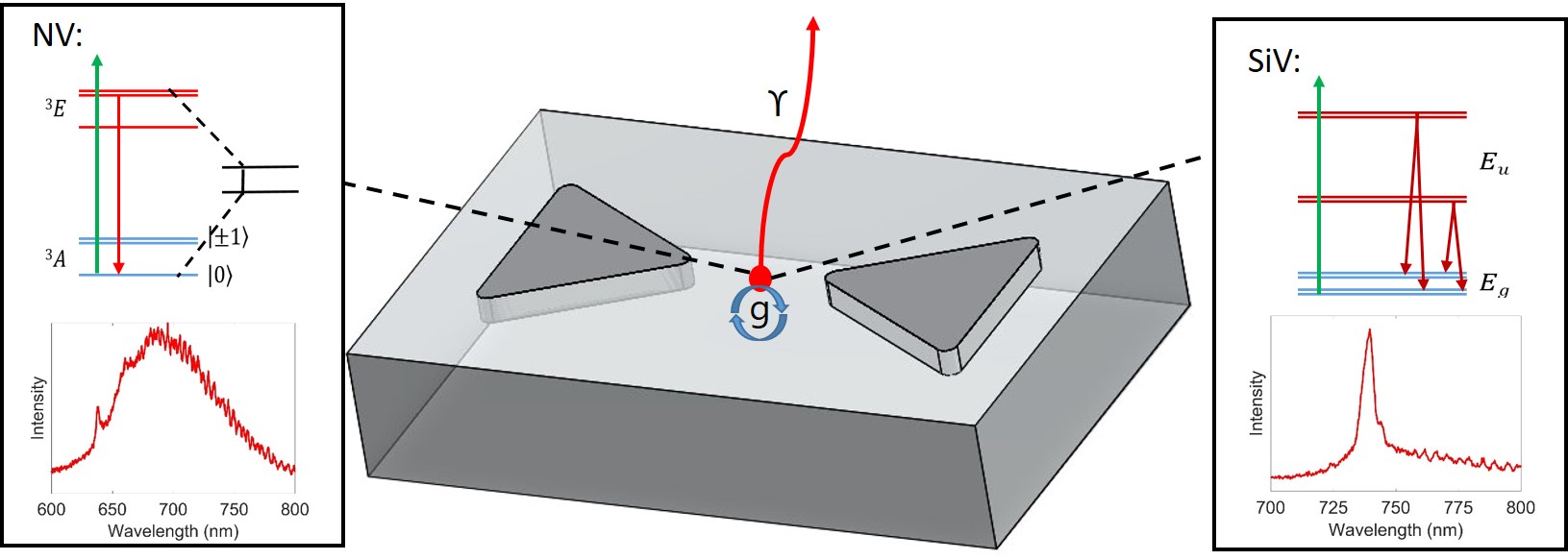}
	\end{minipage}
	\hspace{0cm}
	\caption{A quantum emitter coupled to a metallic nano-antenna.  Left: the energy levels of the NV center and its broad emission spectrum. Right: the energy levels of the SiV defect and its narrow emission spectrum.}
	\label{fig:spectrum}
\end{figure}

\section{Designs}

\subsection{Metallic Structures}
\begin{figure}[h!]
	\begin{minipage}[b]{0.5\linewidth}
		\centering
		\includegraphics[width=7cm]{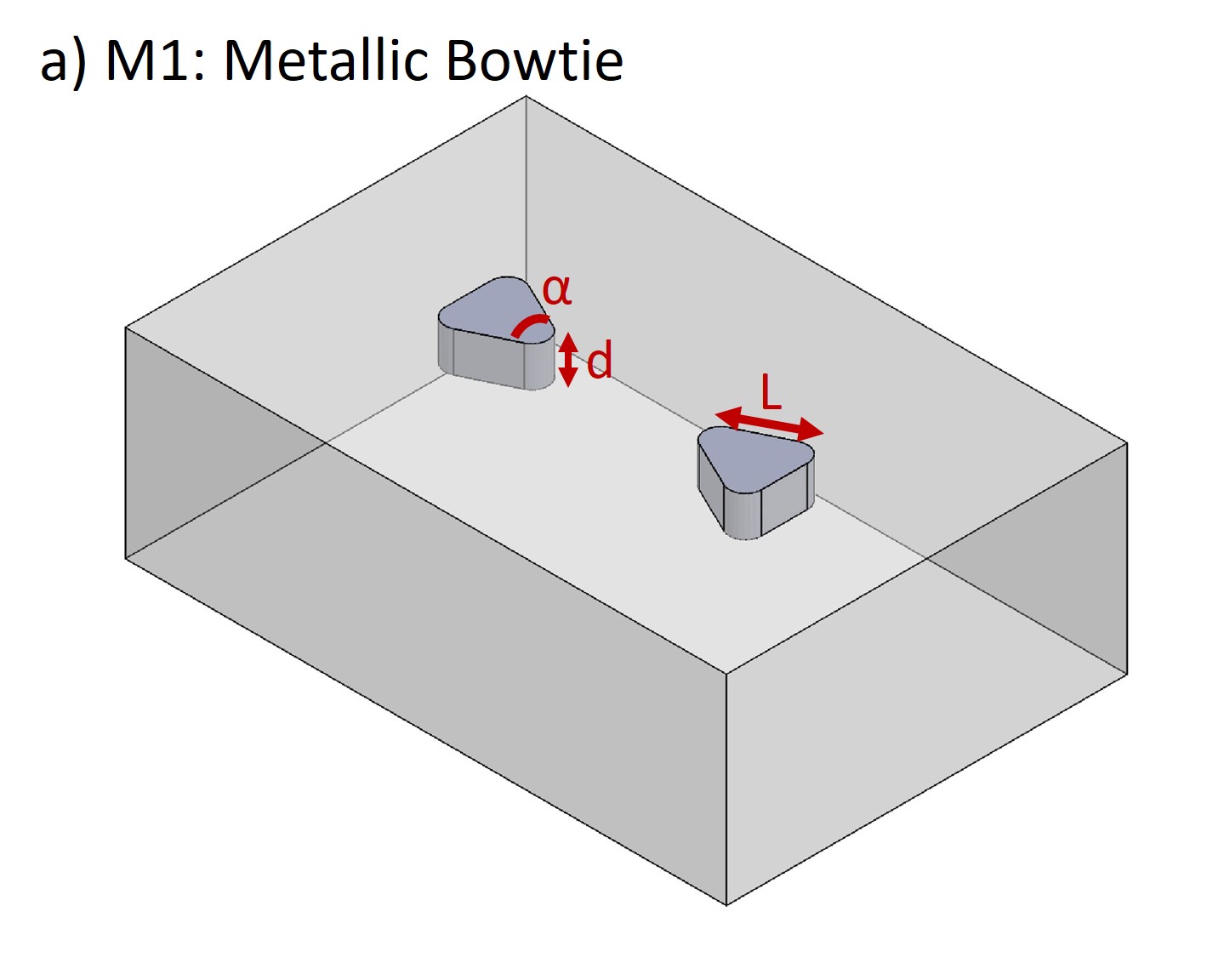}
	\end{minipage}
	\hspace{0cm}
	\begin{minipage}[b]{0.5\linewidth}
		\centering
		\includegraphics[width=7cm]{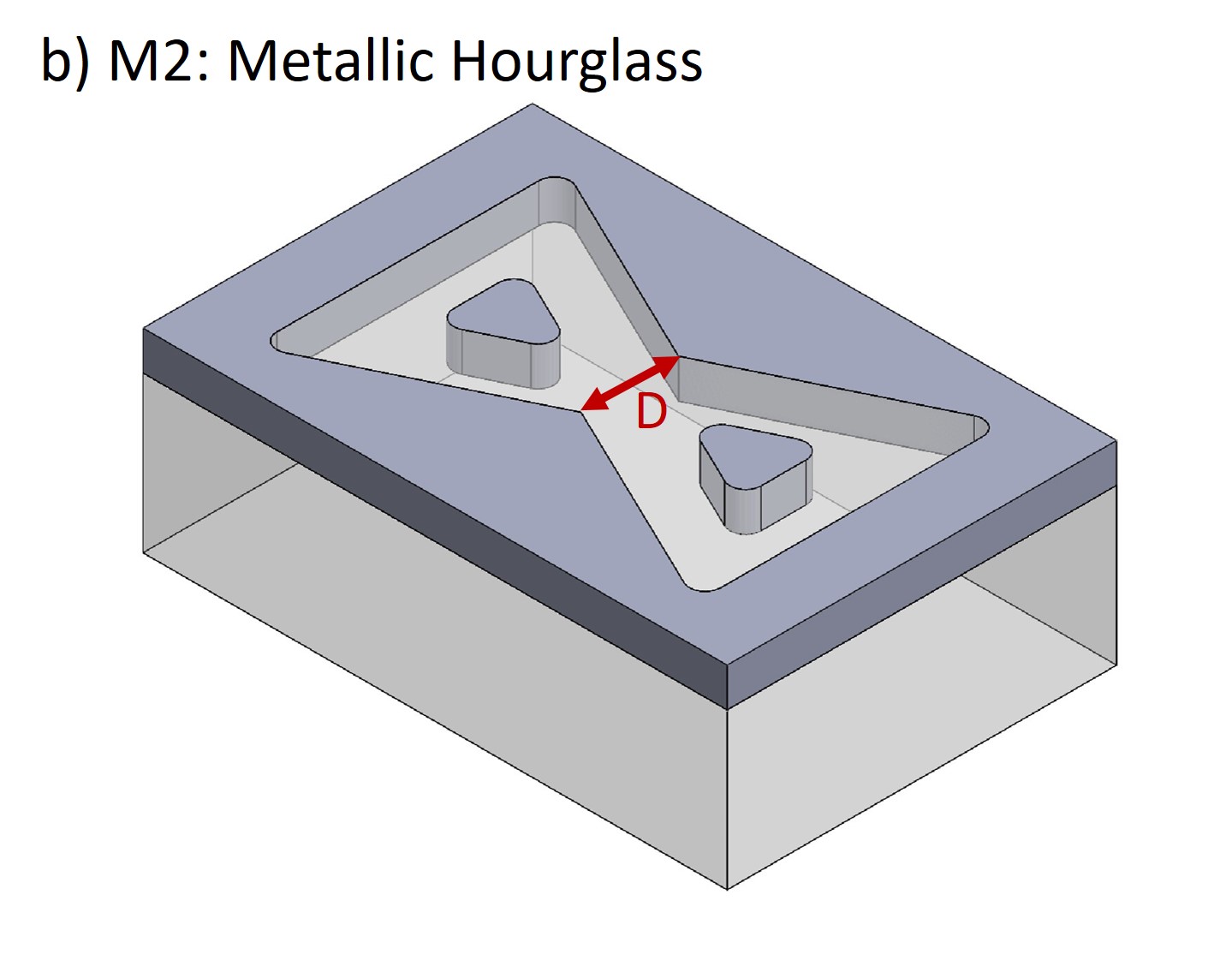}
	\end{minipage}
    \begin{minipage}[b]{0.5\linewidth}
		\centering
		\includegraphics[width=7cm]{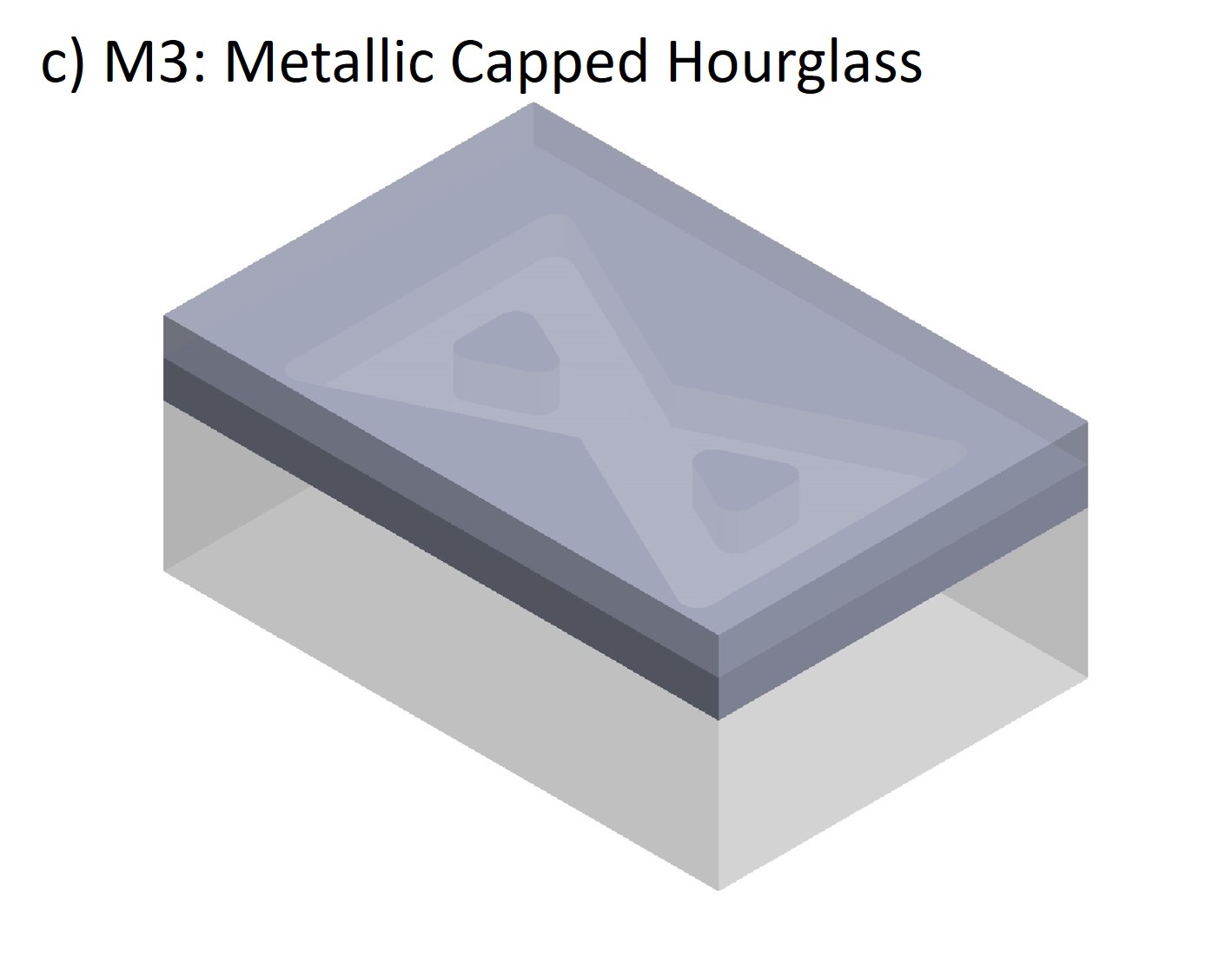}
	\end{minipage}
	\begin{minipage}[b]{0.5\linewidth}
		\centering
		\includegraphics[width=7cm]{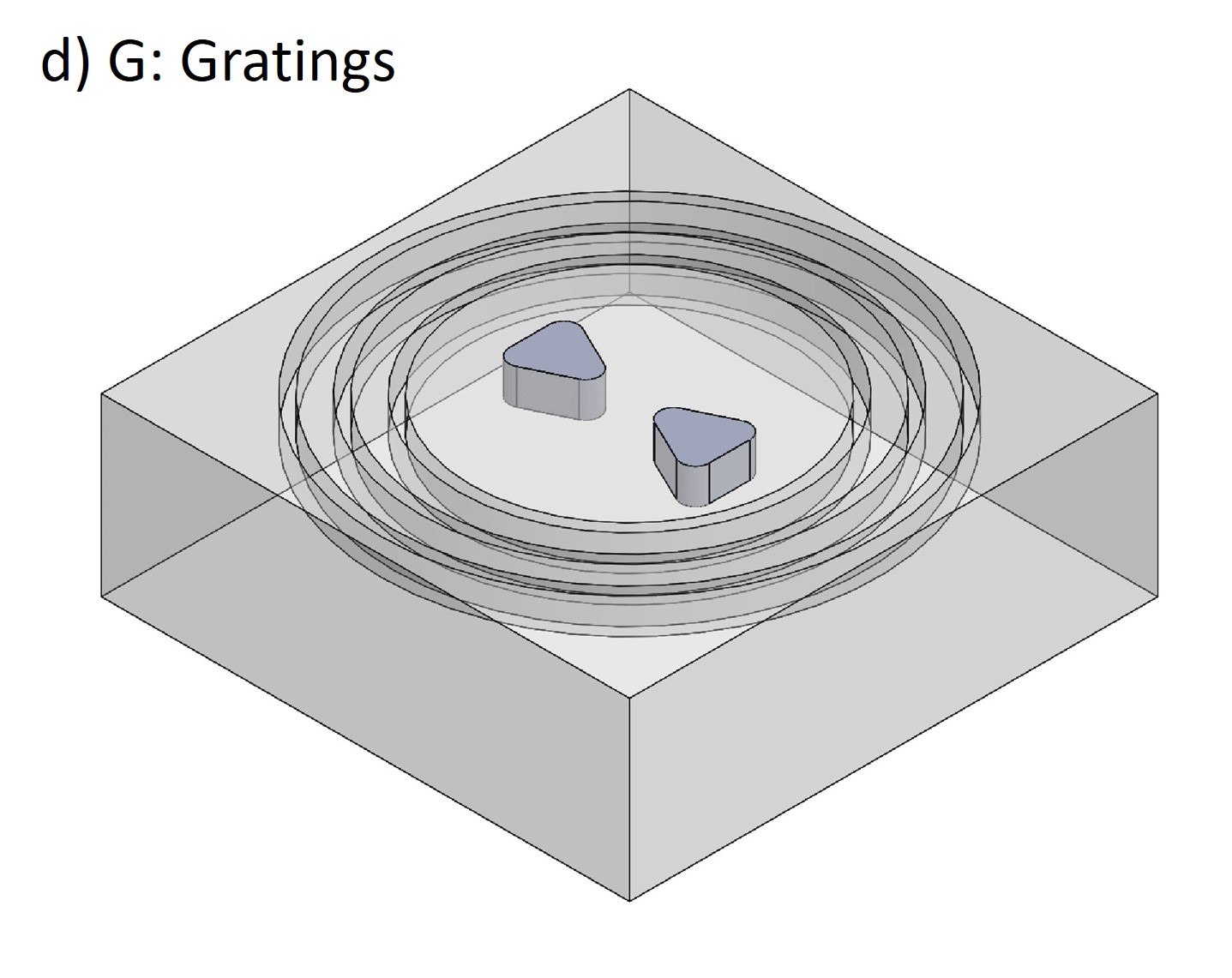}
	\end{minipage}
	\caption{Diamond-embedded metal antennas. (a) M1: a silver bowtie antenna embedded in a diamond medium with length $L$, thickness $d$ and angle $\alpha$ (b) M2: an hourglass nanoantenna consisting of a silver bowtie antenna surrounded by an similar metallic conducting plane with a resulting diamond aperture size $D$. c) M3: a capped hourglass antenna consisting of the hourglass with an additional conformal silver capping layer (d) G-series variants: concentric gratings surrounding the antenna structures are added to increase CE. The labeled parameters were allowed to vary during the optimization.}
    
\label{fig:met_struc}
\end{figure}

Our basic metallic design consists of an embedded bowtie antenna aligned to the color center. Similar designs have been used to couple to quantum emitters in other material systems\cite{2009.Nature.Moerner-Anika,Schuller2010}. As shown in Fig. \ref{fig:met_struc}a, by embedding the bowtie into the plane of the color center (antenna design M1), it is possible to maximize the overlap between the electromagnetic mode and the emitter dipole. To optimize near-field coupling strength and far-field radiation directivity, we expand this basic design to include an additional in-plane conductor, as illustrated in the `hourglass antenna' design of Fig. \ref{fig:met_struc}b (designated M2). Design M3 caps M2 with a conducting half-plane, which redirects any upward radiation down through the diamond into the collection optics, allowing further CE improvement. Finally, we also consider variants M1G, M2G and M3G which add an additional concentric in-plane grating, which has been shown to increase far-field outcoupling \cite{Choy2013} (Fig. \ref{fig:met_struc}c). 

\subsection{Dielectric Structures}

\begin{figure}[h!]
	\begin{minipage}[b]{0.5\linewidth}
		\centering
		\includegraphics[width=7cm]{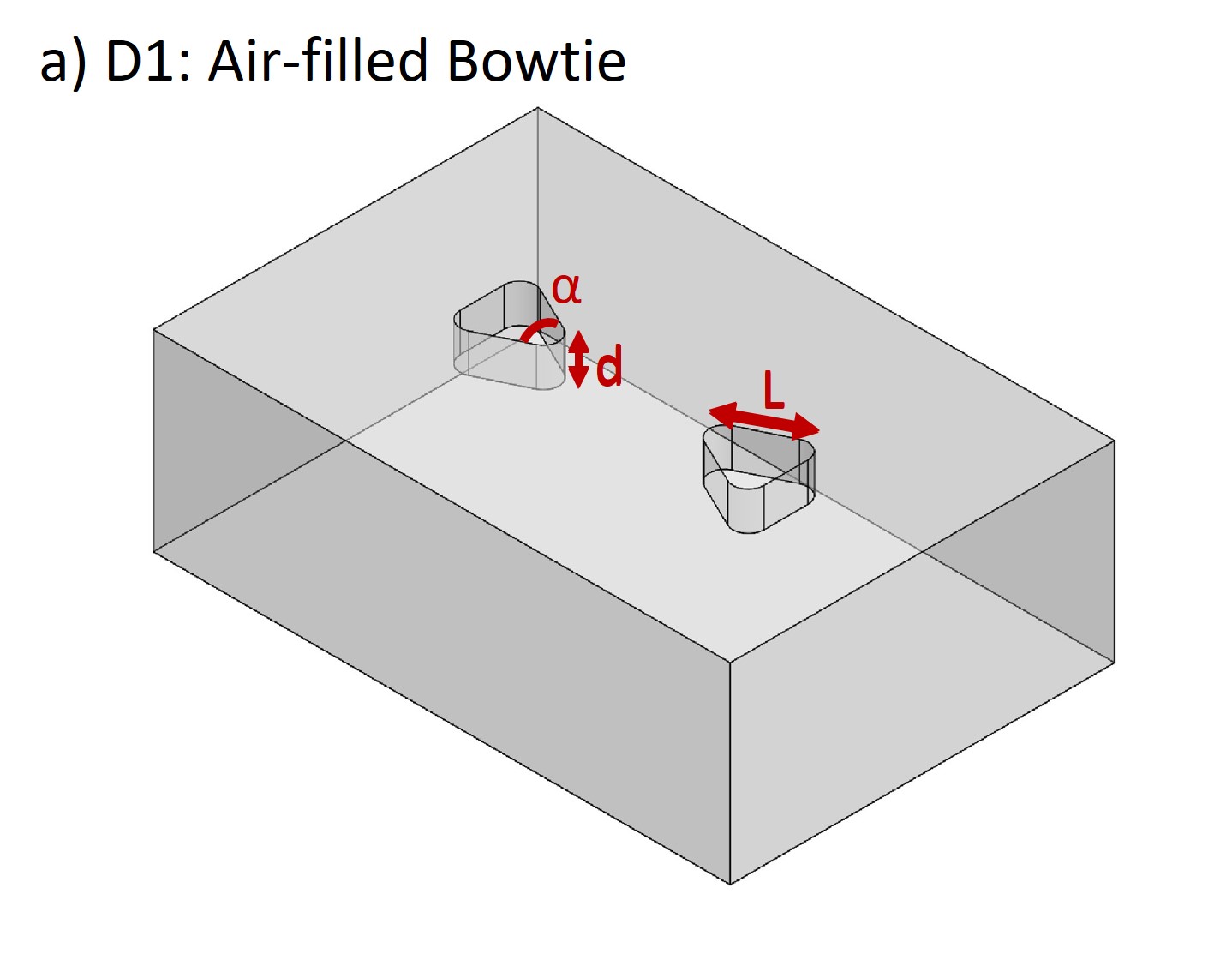}
	\end{minipage}
	\hspace{0cm}
	\begin{minipage}[b]{0.5\linewidth}
		\centering
		\includegraphics[width=7cm]{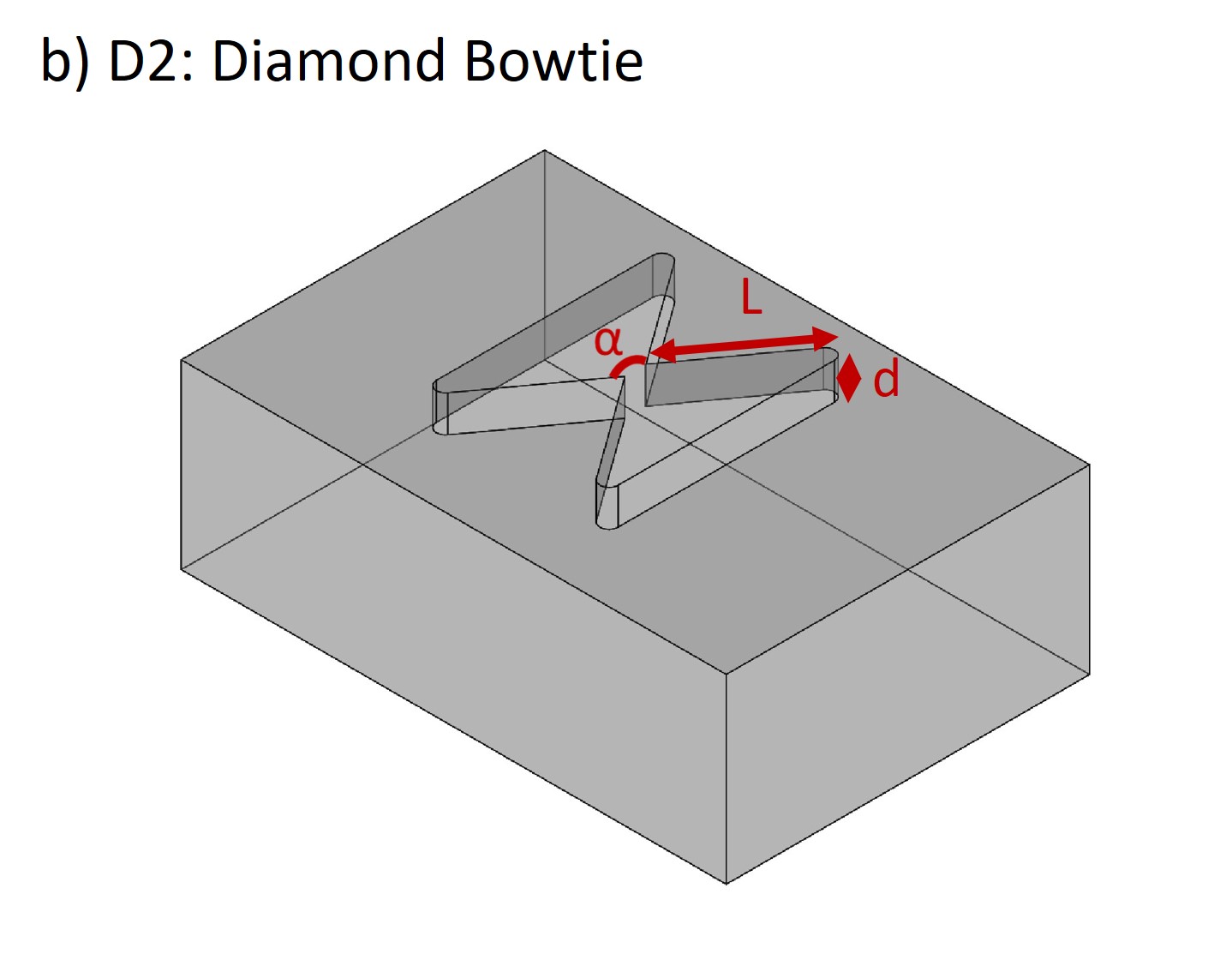}
	\end{minipage}
	\caption{Dielectric antenna designs. (a) D1: a pair of air-filled bowtie tips etched into the diamond with length $L$, thickness $d$ and angle $\alpha$. D1G adds an additional concentric grating similar to Fig. \ref{fig:met_struc}d. (b) D2: raised diamond bowtie with length $L$, angle $\alpha$ and thickness $d$. The labeled parameters were allowed to vary during the optimization.}
    
\label{fig:di_struc}
\end{figure}

Secondly, we consider a class of all-dielectric antenna designs. Dielectric antennas avoid any metal ohmic losses or quenching and can therefore potentially reach high collection efficiencies, and can achieve small mode volumes\cite{Choi2017} resulting in Purcell enhancement . Our first design (Fig. \ref{fig:di_struc}a) consists of a pair of air-filled bowtie tips etched into the diamond with (D1G) and without (D1) bullseye gratings. The second design (D2) is a raised diamond bowtie where the material surrounding the bowtie has been etched away (Fig. \ref{fig:di_struc}b).

\subsection{Hybrid Structures}

\begin{figure}[h!]
	\begin{minipage}[b]{0.5\linewidth}
		\centering
		\includegraphics[width=7cm]{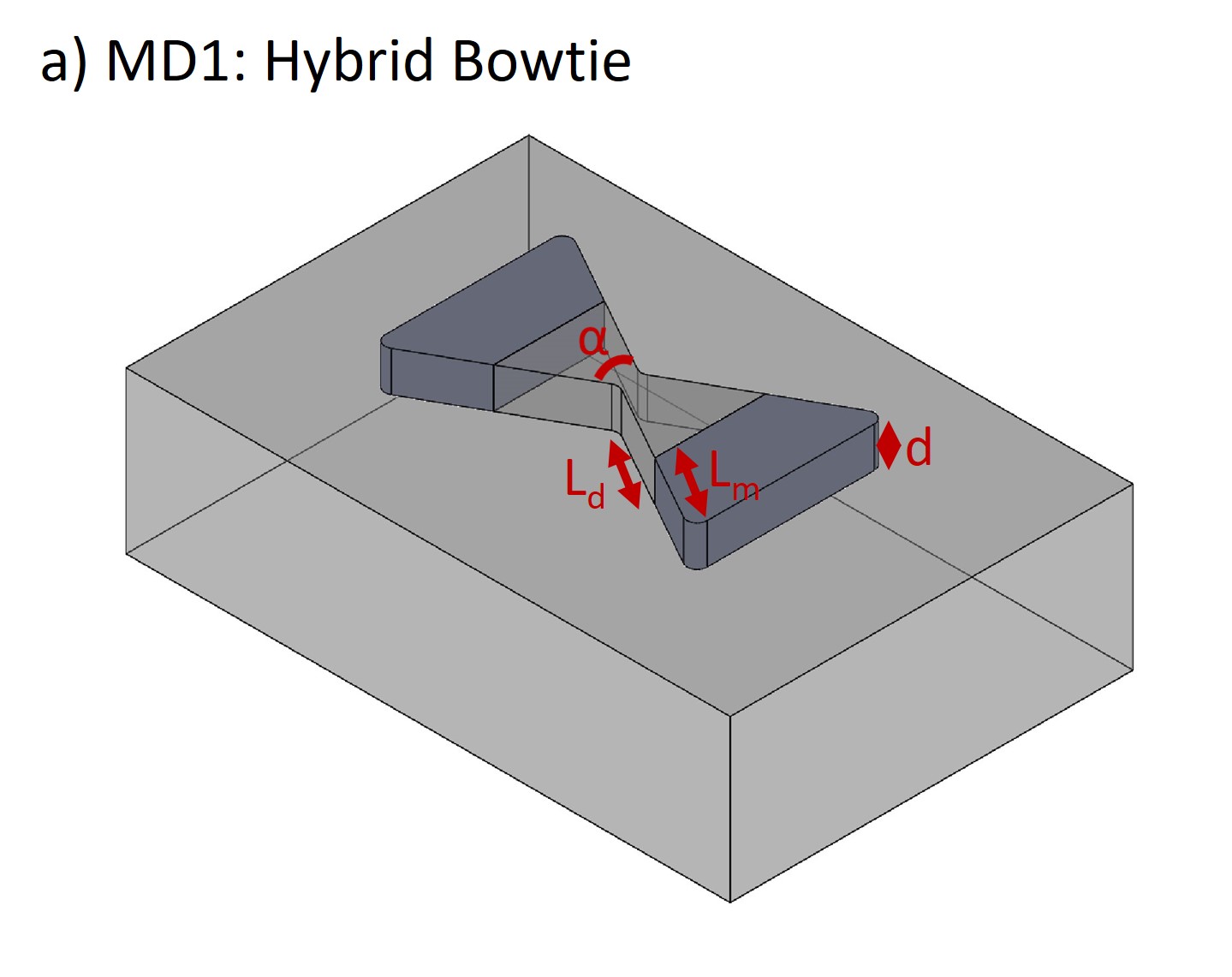}
	\end{minipage}
	\hspace{0cm}
	\begin{minipage}[b]{0.5\linewidth}
		\centering
		\includegraphics[width=7cm]{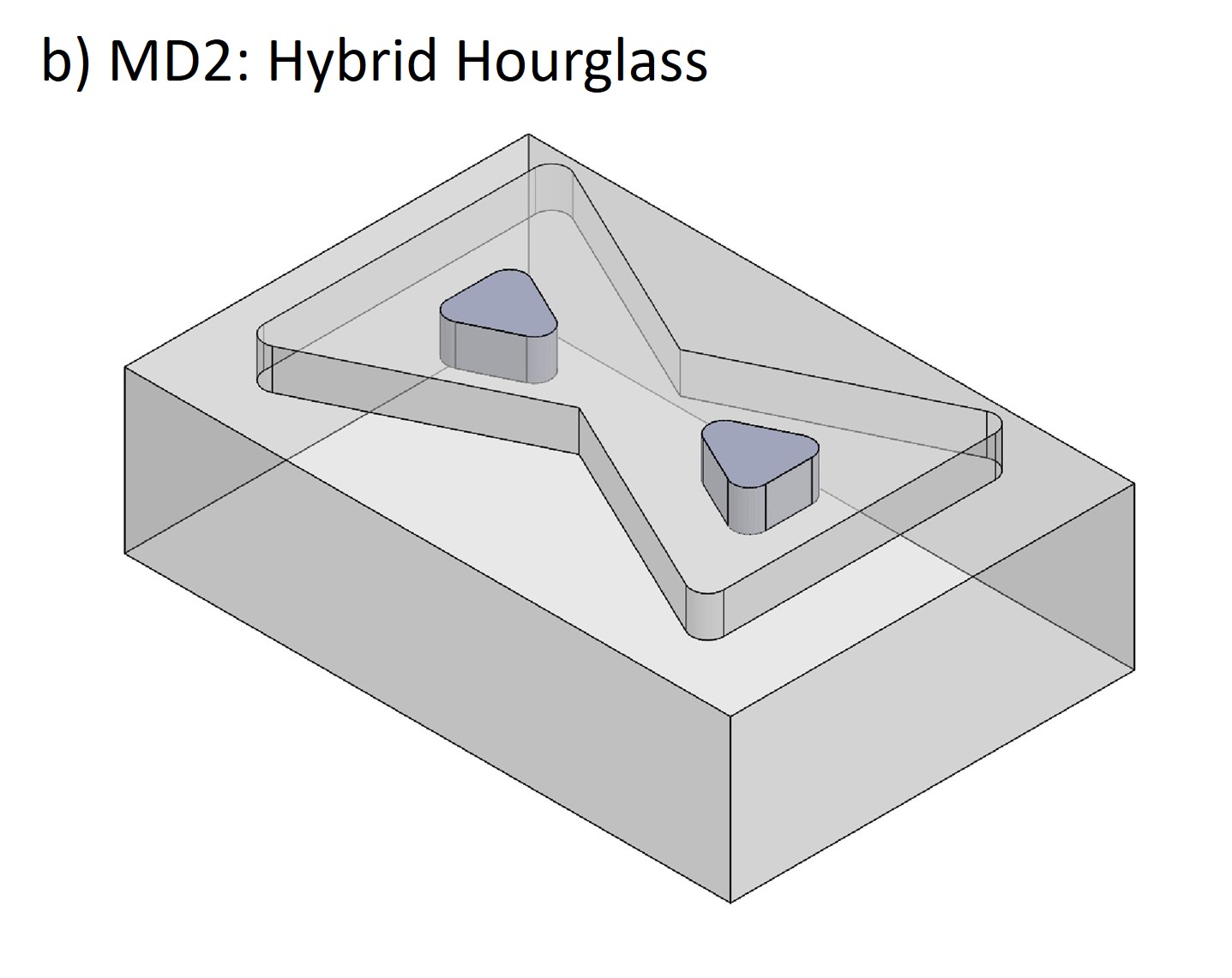}
	\end{minipage}
	\caption{Metal-dielectric antenna designs. (a) MD1: a hybrid bowtie structure with tip angle $\alpha$, dielectric length $L_d$, metal length $L_m$ and thickness $d$. D1G adds an additional concentric grating similar to Fig. \ref{fig:met_struc}d. (b) MD2: a hybrid hourglass structure similar to the metallic hourglass with the same optimization variables.}
    
\label{fig:hybrid_struc}
\end{figure}

The final class of designs considered use patterning of both metallic and dielectric features. These metal-dielectric antennas allow us balance the benefits of the metallic antenna (small mode confinement) with that of the dielectric antenna (low ohmic losses and no quenching). We consider two specific metal-dielectric hybrids. The first (MD1) consists of pair of hybridized bowties with a trapezoidal metallic base, a dielectric (diamond) tip and bridge, and air on the sides of the emitter (Fig. \ref{fig:hybrid_struc}a). We also consider a variant with bullseye gratings, MD1G. The second design (MD2) is a variant of the hourglass structure, with air replacing metal in the similar outer layer (Fig. \ref{fig:hybrid_struc}b). 

\section{Simulation}
We simulated our structures using finite difference time domain methods in the commercial software package Lumerical. Each structure is simulated with a dipole source located at the center of the design and oriented in the plane of the inner bowtie. We used silver as our metal due to its low optical absorbance across the NV emission spectrum. For all metallic structures, we take the distance between the silver bowtie tips to be 20 nm to avoid quenching, and vary the length, thickness, tip angle and aperture width as indicated in \ref{fig:met_struc}. The dielectric and hybrid designs, which avoid quenching, allow us to move the tips closer; this is connected by a bridge of 5 nm. All corners of the nanostructures were smoothed with an 8 nm radius of curvature to account for fabrication limitations Some effects such as the anomalous skin effect are not reflected by our simulations as they should be negligible at 20 nm gap sizes in the visible spectrum \cite{Eggleston2015}.  

For each geometry, we performed an optimization over the degrees of freedom indicated in Figures \ref{fig:met_struc}, \ref{fig:di_struc}, and \ref{fig:hybrid_struc}), as well the dipole depth beneath the diamond surface, to maximize the CPR inside an NA of 0.95 at 650 nm as the figure of merit. We measured the CE both from the top and the bottom of the structure and used the greater of the two for calculating the figure of merit.

The optimal free parameters (the parameters maximizing the CPR) for all antenna classes are summarized in \tablename{2} of the supplementary information. 

\section{Results}

\begin{table}[h!]
	\begin{center}
		\caption{Figures of Merit of Optimized Structures}
		\label{tab:table1}
		\begin{tabular}{|l||c|c|c|c|}
			\hline
			Design & Avg. CPR & CE & F$_P$ \\
			\hlineB{3}
			M1G: Metallic Bowtie w/grating & 12.1 & 6.2\% & 186 \\
			\hline
			M2G: Metallic Hourglass w/grating& 12.8  & 8.4\% & 164 \\
			\hline
			M3G: Metallic Capped Hourglass w/grating& 17.5  & 8.5\% & 180 \\
			\hlineB{3}
			D1G: Air-filled Bowtie w/grating & 3.3  & 30\% & 11.2 \\
			\hline
            D2: Diamond Bowtie  & 6.8  & 70\% & 9.6 \\
			\hlineB{3}
            MD1G: Hybrid Bowtie w/grating & 13.4  & 11\% & 112 \\
			\hline
            MD2: Hybrid Hourglass  & 25.6  & 13\% & 199 \\
			\hline
		\end{tabular}
	\end{center}
\end{table}

\subsection{Metallic Structures}
\begin{figure}[h!]
	\centering 
	\includegraphics[width=14cm]{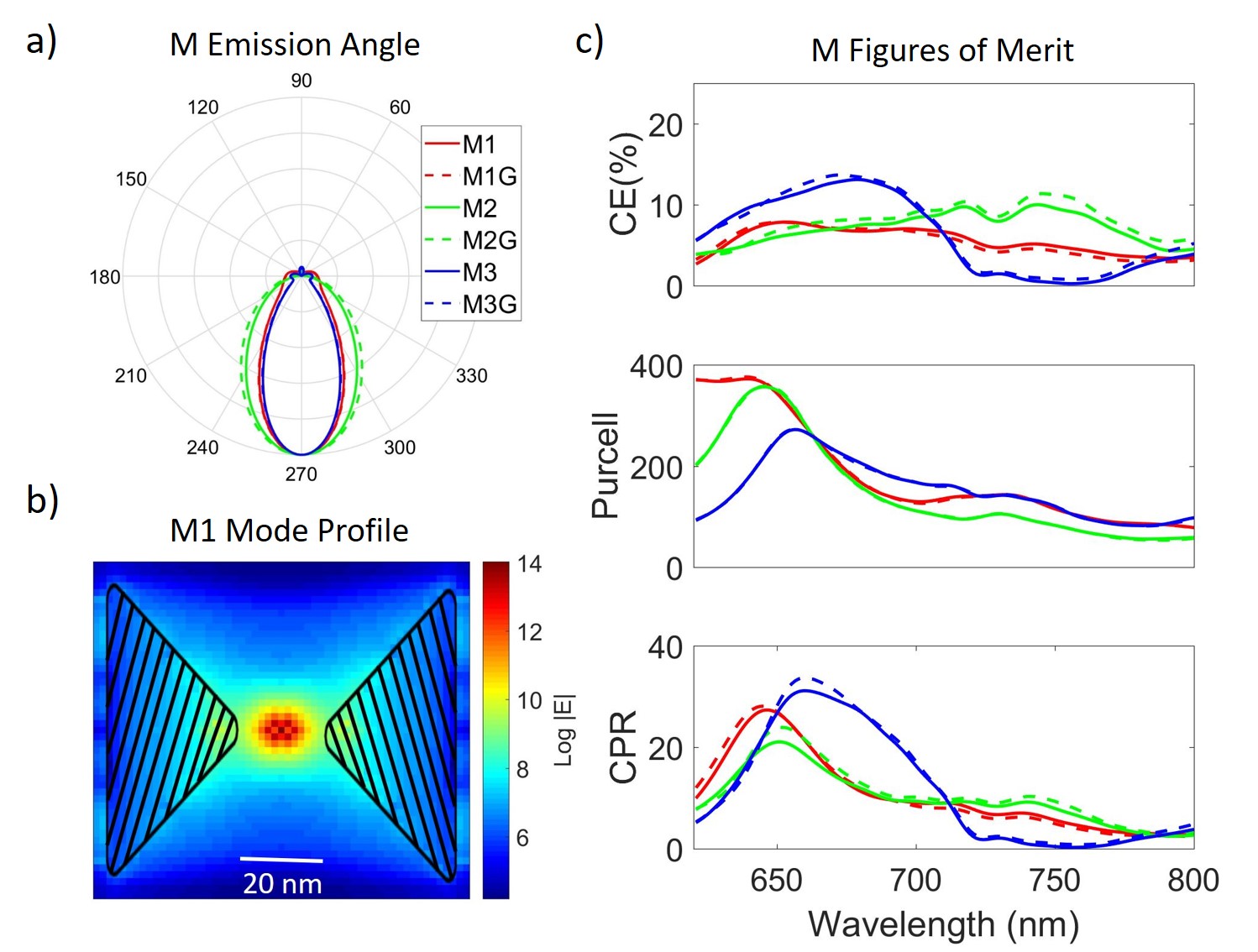}
	
	\caption{(a) The emission pattern for each of the metallic structures. All structures demonstrate a directional emission pattern (b) The mode profile of the electric field confined by the metallic nano-antennas. The structures demonstrate a small mode volume. (c) The figures of merit of each metallic designs over the NV spectrum.}
	\label{fig:metallic}
\end{figure}

The CE depends on the directionality of emission and any optical losses inside the antenna. The farfield emission patterns of the metallic structures are shown in Fig. \ref{fig:metallic}a. All three optimized designs achieve directional emission into small angles: 75$\%$ of the farfield emission is collected between an NA of 0.5 and 0.7 for all designs. Interestingly, the addition of grating couplers (M1-M3G) does not significantly improve the directionality of our designs when compared to that of the optical nanoantennas themselves. The capped design M3 emits into the smallest half-angle, with 90$\%$ of the emission collected within NA of 0.95 (Fig. \ref{fig:metallic}c). While all designs enhance the CE for every wavelength in the spectrum of the NV, the capped M3 design achieves the highest CE, with a threefold increase in the region where the NV spectrum is maximal. While these results show the strong directionality of these structures, the maximal collection efficiency is limited to a fraction (12$\%$) of the overall emission even with a unity-NA collection aperture. The drop in CE is primarily due to ohmic loss in the antenna, which results in a loss of 75$\%$ within the nanostructure itself, and secondarily due to total internal reflection at the diamond-air interface below the nanostructure which results in a loss of 13$\%$.

The second element of the figure of merit is the Purcell enhancement of the emitter. Due to the small mode volumes of these structures (Fig. \ref{fig:metallic}b), each plasmonic nanoantenna achieves high Purcell enhancement. The spectrally resolved Purcell factor between 620 nm to 800nm wavelength can be seen in Fig. \ref{fig:metallic}c. This broadband enhancement indicates that these structures are not high quality resonators but rather efficient antennas with their quality factor Q$\approx$ 15 dominated by the ohmic loss in the silver. The design consisting only of bowtie antennas offers the highest peak Purcell factor, however, the enhancement significantly varies across the spectrum. Unlike the bowtie antenna, the Purcell enhancement from the capped hourglass design has low spectral dependence which makes it desirable for broadband operation.

The final figure of merit, collected photon rate, is shown in Fig. \ref{fig:metallic}c. Due to the broadband nature of the antennas, we achieve a high CPR across the spectrum from all designs, and in particular by the capped hourglass design. We then computed the figure of merit normalized to the NV fluorescence spectrum for each design. Here we see that the capped design with gratings achieves the highest overall collected photon rate of 17.51 averaged through out the spectrum. By comparison, a bare emitter in bulk diamond has a CPR of $\approx$ 0.06 averaged over the spectrum, with a Purcell factor of $\approx 1$ and a wavelength dependent CE ranging between $6\%-8\%$. The optical nanoantennas designed by our method improve upon this by a factor of 250, well within the range required for single-shot spin readout at room temperature.

% \begin{figure}[h!]
% 	\hspace{0cm}
	
% 	\begin{minipage}[b]{0.5\linewidth}
% 		\centering
% 		\includegraphics[width=7cm]{CE.jpg}
% 		\label{CE}
% 	\end{minipage}
% 	\begin{minipage}[b]{0.5\linewidth}
% 		\centering
% 		\includegraphics[width=7cm]{Purcell.jpg}
% 		\label{Purcell}
% 	\end{minipage}
% 	\begin{minipage}[b]{0.5\linewidth}
% 		\centering
% 		\includegraphics[width=7.5cm]{CExPurcell.jpg}
% 		\label{PC}
% 	\end{minipage}
% 	\caption{Performance of proposed designs over the NV spectrum. (a) shows wavelength dependence of the Purcell factor of all designs. In (b) we see the wavelength dependence of the collection efficiency. (c) shows the Purcell-CE product as a function of wavelength throughout the NV spectrum.}
% 	\label{fig:plots}
% \end{figure}

\subsection{Dielectric Structures}

\begin{figure}[h!]
	\centering 
	\includegraphics[width=14cm]{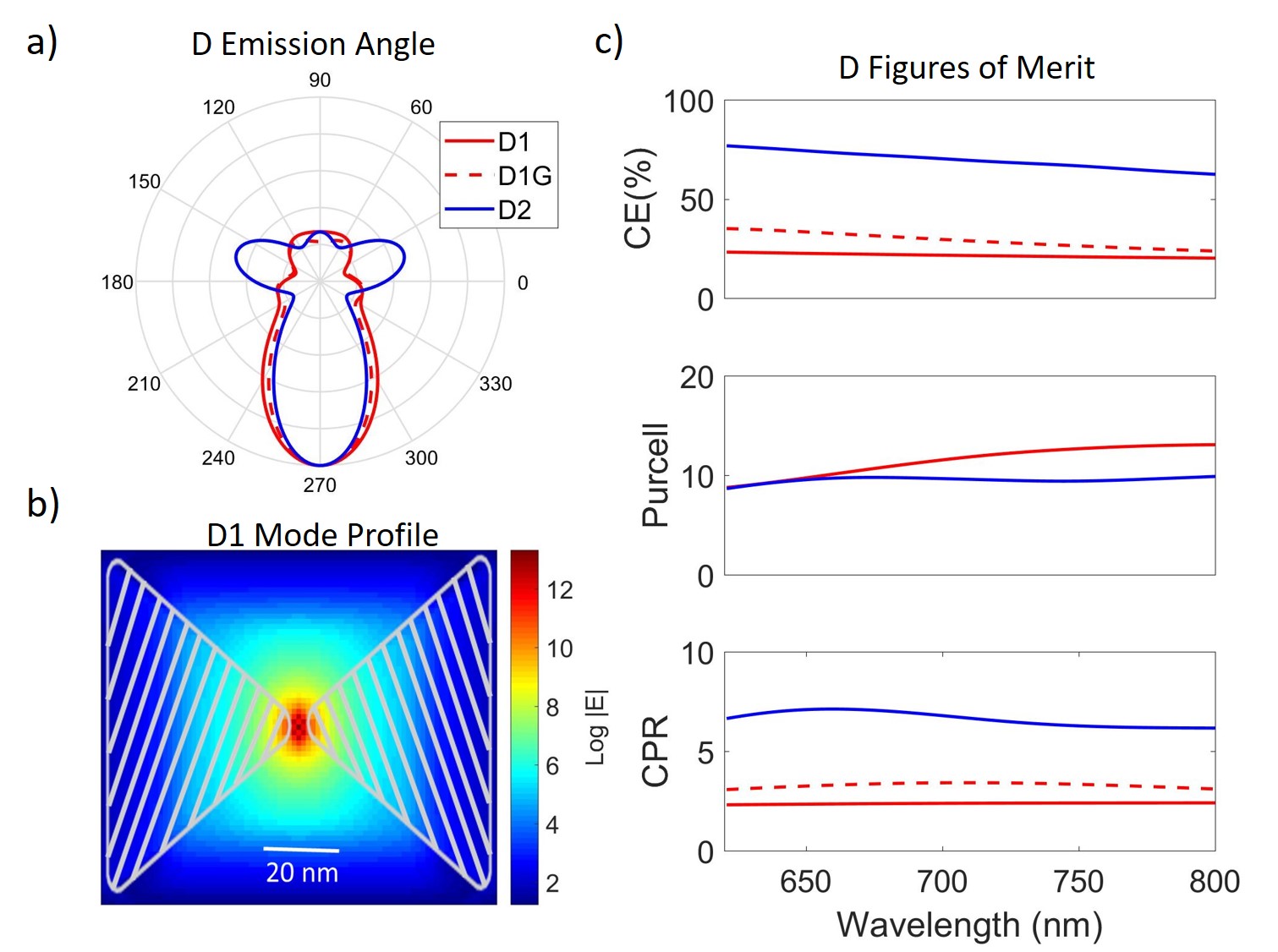}
	
	\caption{(a) The emission pattern for each of the optimized dielectric structures. (b) The mode profile of the electric field confined by the dielectric nanoantennas. The air tips confine the electric field into a small mode volume. (c) The figures of merit of each optimized dielectric designs through the NV spectrum.}
	\label{fig:dielectric}
\end{figure}

The optimized figures of merit for the dielectric structures are shown table 1, with spectrally-resolved data shown in Fig. \ref{fig:dielectric}. We observe that the air-filled bowtie (D1) demonstrates a directional emission pattern (Figure \ref{fig:di_struc}a). Unlike for the metallic antennas, the addition of circular dielectric gratings (D1G) further tightens the emission pattern. The optimized D2 design (consisting of the raised diamond bowtie), has lower directionality. While the D2 shows less directionality in the far-field, it has a much better CE (Figure \ref{fig:dielectric}c) due to lower loss from total internal reflection at the bottom air-diamond interface.

As can be seen in Figure \ref{fig:dielectric}b, the dielectric antennas achieve a small mode volume and enhance the spontaneous emission rate of the NV. We observe a Purcell factor varying between 8 and 13 across the NV spectrum for both designs. While this figure is an order of magnitude smaller than the enhancement provided by metallic antennas, our dielectric designs do not suffer from non-radiative quenching and loss caused by proximity to the antenna tips, resulting in a higher quantum efficiency. The final full-spectral CPR for D1 and D2 of 3.3 and 6.8, respectively, reflects this low Purcell enhancement with high collection efficiency.

\subsection{Hybrid Structures}

\begin{figure}[h!]
	\centering 
	\includegraphics[width=14cm]{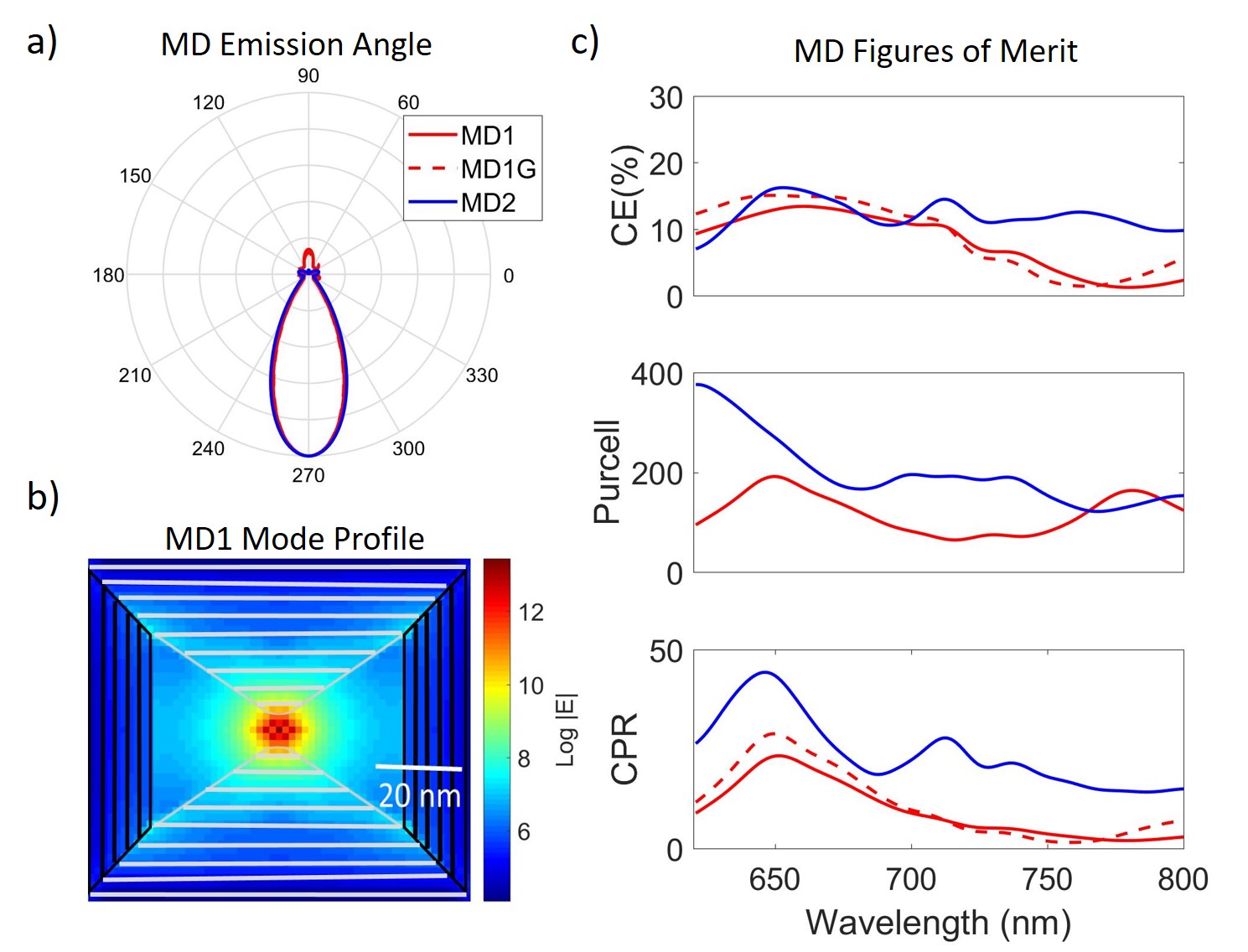}
	
	\caption{(a) The emission pattern for each of the optimized hybrid structures. (b) The mode profile of the electric field confined by the hybrid nano-antennas. The air-tips and metal trapezoids confine the electric field into a small mode volume. (c) The figures of merit of th each hybrid designs through the NV spectrum.}
	\label{fig:hybrid}
\end{figure}

Both of our hybrid designs demonstrate directional emission into a small NA (Figure \ref{fig:hybrid}a), with the hybrid hourglass MD2 achieving a higher degree of directionality downward into the collection optics as compared to MD1 which has significant emission upward and out of the diamond. The hybrid hourglass structure MD2 offers broad increase in CE across the NV spectrum, increasing the CE three-fold as compared to a bare dipole. The hybrid bowtie MD1 also shows an improvement in CE, which is further enhanced by the addition of bullseye gratings. The CE improvement of these structures over the metallic designs is due to the lower ohmic loss of the hybrid structures (50\% of total emission) as compared to the metallic (75\%).

Both designs show a Purcell enhancement of two orders of magnitude across the NV spectrum. In the hybrid hourglass structure this enhancement is due to the plasmonic antennas surrounding the emitter, whereas in the hybrid bowtie structure both the air tips and plasmonic metal contribute to the Purcell factor. We can see that both designs offer a small mode volume (\ref{fig:hybrid}b), which directly contributes to the Purcell enhancement. The combination of large Purcell enhancement with improved collection efficiency as compared to metallic designs results in a high CPR of 25.6 for the hybrid bowtie. 

\section{Discussion}
While the metallic designs offer a large Purcell enhancement, the CE remains relatively low due to coupling non-radiative modes. To overcome these losses we proposed and studied dielectric and hybrid structures, which improve the CE largely due to the elimination of ohmic loss in the metallic tips. In particular, the hybrid structures produce Purcell enhancement comparable to the metallic designs, while reducing the coupling to non-radiative modes.

There are several avenues to further increase the CPR. One is through relaxing the fabrication-imposed constraints on gap size and radius of curvature. In this work, we set a lower bound of 20 nm for the distance between metallic tips, 5 nm for the air tips, and limit the radius of curvature for any corner to 8 nm. These fixed constraints can significantly affect the CPR, with smaller distances resulting in greater CPR. However, in designs using metals the smaller gap distance will decrease quantum efficiency and CPR. Non-local effects can also become significant for smaller gap sizes, which quench emission and are difficult to simulate\cite{Eggleston2015}. Furthermore, it has been shown that proximity to metallic surfaces can result in charge instability and spin depolarization for diamond quantum emitters\cite{Kolkowitz}. The hybrid bowtie designs could alleviate these issues; however, we have fixed the gap size for an equivalent basis of comparison.

Finally, these structures can be used to enhance the CPR of other diamond quantum emitters, such as silicon-vacancy and germanium-vacancy centers. These emitters are not as broadband as NV centers, and so higher enhancements can be obtained within their narrower spectral band. For silicon-vacancy centers, a metallic bowtie can enhance the CPR to 35.7, a factor of two improvement over the NV enhancement achieved with the same design.

\section{Conclusion}

This work optimized metal-, dielectric-, and metal-dielectric antenna designs for diamond quantum emitters. Numerical simulations showed a spontaneous emission rate and CE enhancement over the full NV spectrum of 25.6, three orders of magnitude greater than the detection rate under standard oil-immersion microscopy ($\sim$ 0.06), and exceeding by an order of magnitude the collected photon rate even with perfect collection efficiency. The hybrid metal-dielectric antenna could also be of interest for plasmon-enhanced light emitting diodes, for which CPR is a central figure of merit\cite{2000.IEEE.Vuckovic.plasmon_LED,Eggleston2015}. For quantum emitters, the proposed antenna designs promise efficient spin-photon interfaces that are important for applications in quantum sensing and quantum information processing. 

\section{Supplement}
\begin{table}[h!]
	\begin{center}
		\caption{Metallic optimized parameter values}
		\label{tab:table1}
		\begin{tabular}{|c||c|c|c|c|}
			\hline
			Design & $\alpha$ & $L$ & $d$ & $D$\\
			\hline
			M1: NV Bowtie Antenna & 109$^\circ$ & 55nm & 139nm & NA\\
			\hline
			M2: NV Hourglass & 140$^\circ$ & 189nm & 108nm & 400nm \\
			\hline
			M3: NV Capped Hourglass & 138$^\circ$ & 99nm & 74nm & 414nm \\
			\hline
			M3: SiV Capped Hourglass & 128$^\circ$ & 120nm & 97nm & 500nm \\
			\hline
		\end{tabular}
	\end{center}
\end{table}

\begin{table}[h!]
	\begin{center}
		\caption{Dielectric optimized parameter values}
		\label{tab:table2}
		\begin{tabular}{|c||c|c|c|}
			\hline
			Design & $\alpha$ & $L$ & $d$ \\
			\hline
			D1: NV Dielectric Bowtie & 72$^\circ$ & 300nm & 119nm \\
			\hline
			D2: NV Dielectric Antenna & 155$^\circ$ & 775nm & 311nm  \\
			
			\hline
		\end{tabular}
	\end{center}
\end{table}

\begin{table}[h!]
	\begin{center}
		\caption{Hybrid optimized parameter values}
		\label{tab:table3}
		\begin{tabular}{|c||c|c|c|c|}
			\hline
			Design & $\alpha$ & $L$ & $d$ & $D$\\
			\hline
			NV Hybrid Hourglass & 91$^\circ$ & 50nm & 250nm &400nm \\
			\hline
            
		\end{tabular}
        
        \begin{tabular}{|c||c|c|c|c|}
			\hline
			Design & $\alpha$ & $L_d$ & $L_m$ & $d$\\
			\hline
			NV Hybrid Bowtie & 71$^\circ$ & 25nm & 75nm &68nm \\
			\hline
            
		\end{tabular}
	\end{center}
\end{table}

\section*{Acknowledgments}
The authors would like to thank Tim Schr{\"o}der and Fr{\'e}d{\'e}ric Peyskens for helpful discussions, and Megan Yamoah for graphic design assistance. A.K. would also like to thank the MIT EECS - Lincoln Labs Undergraduate Research and Innovation fund for their financial support during this research. M.T. acknowledges support from the Army Research Office Multidisciplinary University Research Initiative (ARO MURI) biological transduction program. D.E. acknowledges support from the Center for Integrated Quantum Materials (NSF grant DMR- 1231319) and the NSF program ACQUIRE:``Scalable Quantum Communications with Error-Corrected Semiconductor Qubits.

\end{document}